\documentclass[prb,aps,amssymb,twocolumn,showpacs,floatfix]{revtex4}
\usepackage[dvips]{graphicx}
\def\be{\begin{equation}}
\def\ee{\end{equation}}
\def\bea{\begin{eqnarray}}
\def\eea{\end{eqnarray}}

\begin{document}

\title{Quantum Hall ferromagnets, cooperative transport anisotropy,\\
and the random field Ising model}

\author{J.T. Chalker$^{1,2}$}
\author{D. G. Polyakov$^{3}$}\altaffiliation[Also at ]{Also at 
A. F. Ioffe Physico-Technical Institute, 194021 
St. Petersburg, Russia.}
\author{F. Evers$^{3}$} 
\author{A. D. Mirlin$^{1,3}$}\altaffiliation[Also at ]{Also at 
Petersburg Nuclear Physics Institute,
188350, St Petersburg, Russia.} 
\author{P. W\"olfle$^{1,3}$}
\affiliation{\mbox{$^{1}$Institut f\"ur Theorie
der~Kondensierten~Materie,~Universit\"at~Karlsruhe,~76128~Karlsruhe, Germany}\\
$^{2}$Theoretical Physics, University of Oxford,
1 Keble Road, Oxford OX1 3NP, United Kingdom\\
$^{3}$Institut f\"ur Nanotechnologie, Forschungszentrum Karlsruhe,
Karlsruhe 76021, Germany }
\date{\today}

\begin{abstract}
We discuss the behaviour of a quantum Hall system when two
Landau levels with opposite spin and combined filling factor
near unity are brought into energetic coincidence using an
in-plane component of magnetic field. 
We focus on the interpretation of recent experiments under these
conditions [Zeitler {\it et al},
Phys. Rev. Lett. {\bf 86}, 866 (2001); Pan {\it et al}, Phys. Rev. B {\bf 64},
121305 (2001)], in which a large resistance anisotropy develops 
at low temperatures.
Modelling the systems involved as Ising quantum Hall ferromagnets, 
we suggest that this transport anisotropy
reflects domain formation induced by a random field
arising from {\it isotropic} sample surface roughness.
\end{abstract}


\maketitle

Two very striking experimental observations of large electronic transport
anisotropy for quantum Hall systems in tilted magnetic fields have been 
reported recently.\cite{Zeitler,Pan} In both cases, anisotropy appears
at integer values of the filling factor $\nu$
with an in-plane magnetic field component 
tuned to bring two Landau levels of opposite spin into energetic coincidence.
While the in-plane magnetic field component itself defines an axis
within the sample, the fact that large anisotropy appears in
resistivity only below a characteristic temperature of about $1$ kelvin 
suggests it has a cooperative origin. 
Our aim in this paper is to develop a 
theoretical treatment of such systems and to discuss the source 
of the observed anisotropy.

In view of the phenomenological similarities, it is natural 
to make comparisons between these Landau level coincidence
experiments and the earlier discovery of resistance anisotropy in
quantum Hall systems near half-filling of high 
Landau levels,\cite{Eisenstein} attributed to the formation
of a uniaxial charge density wave with a period set by the cyclotron
radius.\cite{CDW} Some distinctions are, however, 
clear. Most importantly, the
nature of the electron states near the chemical potential and their
average occupation is quite different in each case: two separate 
orbital Landau levels with opposite spin and a combined filling
factor close to unity are involved in the coincidence experiments, 
as against a single, spin-polarised and roughly half-filled Landau level 
in the other case. In this context it is desirable to examine
a range of possible explanations for low-temperature anisotropy.

The study of cooperative effects in coincident Landau levels
has a long history. Consider a system with fixed magnetic field
strength $B_{\perp}$ perpendicular to the two-dimensional 
electron or hole gas, as
a function of total field strength $B_{\rm tot}$.  
In a single-particle description there are pairs of Landau levels 
having opposite spin orientations and orbital quantum numbers
differing by $\Delta N$, which are separated in energy by
$\Delta N \hbar \omega_c - g^* \mu_{\rm B} B_{\rm tot}$, 
where $\omega_c \propto B_{\perp}$
is the cyclotron frequency and $g^* \mu_{\rm B} B_{\rm tot}$ 
is the Zeeman contribution.
This energy gap falls to zero at coincidence.
Inclusion of exchange interactions
leads, at a combined filling factor of unity for the crossing levels, to
a first-order transition between two ground states in which one
or other level is completely filled. Within a Hartree-Fock treatment, 
the excitation gap remains non-zero through 
this transition.\cite{Quinn}
Early observations of a quantised plateau in Hall resistivity $\rho_{xy}$ and
deep minimum in diagonal resistivity $\rho_{xx}$, 
both persisting through the
transition,\cite{Koch} provide support for such a picture,
while measurements of the temperature dependence of 
$\rho_{xx}$ allow determination 
of an activation energy gap\cite{Daneshvar} 
which has the variation with $B_{\parallel}$
expected theoretically.
This Landau level coincidence transition is one example
of a broad class of cooperative phenomena in quantum
Hall systems, involving ferromagnetism of either spin or pseudospin
variables, which have been a focus for much recent 
work.\cite{Sondhi,Girvin} 
From such a perspective,  
representing the two Landau levels involved using two states of a pseudospin,
interactions between pseudospins are ferromagnetic with Ising
anisotropy,\cite{Jungwirth} while $B_{\rm tot}$ measured from its
value at coincidence acts on pseudospins as a Zeeman field.

With this background in mind, we return to a discussion of 
transport anisotropy. Following the suggestions of Refs.\,\onlinecite{Zeitler}
and  \onlinecite{Pan}, it is clear that the presence of 
a spin or charge density wave could potentially explain this observation.
From the original Hartree-Fock calculations,\cite{Quinn} 
which were for a one-band model and considered only 
trial states with homogeneous charge density, it was found that
instability to a spin density wave is preempted by
the first-order coincidence transition.
An escape from this conclusion 
might be provided by the fact that
the sample involved is in one experiment 
a Si/GeSi heterostructure,\cite{Zeitler}
and in the other a wide quantum well:\cite{Pan} the former 
has valley degeneracy and the latter has two occupied subbands. 
Alternatively, it might be that a better trial state
in Hartree-Fock theory, or calculations which go beyond this
approximation, would yield
a stripe phase  as the true ground state near coincidence.
However, recent Hartree-Fock calculations 
for bilayer systems\cite{Demler}
give only ferromagnetic pseudospin order with parameters relevant
in the present context, as do calculations for one-band
models, in which Hartree-Fock
solutions with both spin and charge density modulations are 
considered,\cite{Evers}
or the Hamiltonian with realistic interaction potentials
is diagonalised exactly for a small number of electrons.\cite{Rezayi}

The observations of Refs.\,\onlinecite{Zeitler}
and  \onlinecite{Pan} therefore present a puzzle, which we argue 
in the following can be understood in terms of domain formation,
with a characteristic size
much larger than the relevant scale for
stripe phases, the cyclotron radius. 
Our account involves three distinct ingredients.
First, we suggest that domains are induced by a random Zeeman field acting 
on the pseudospins, which arises from the interplay between
isotropic sample surface roughness and the in-plane component of magnetic
field. Second, we show that a random field generated by this mechanism
is intrinsically endowed with anisotropic correlations, and that the
correlation anisotropy is large enough to 
explain the observed anisotropy in resistivity.
Third, we argue that transport in a multi-domain sample occurs
along domains walls, via the processes discussed recently in 
Refs.\,\onlinecite{Falko,Mitra,Brey}.
The onset temperature for transport anisotropy
arising by this mechanism is the Curie temperature of the
Ising quantum Hall ferromagnet, and we note that
the reported\cite{Zeitler,Pan} 
onset temperature of about $1$ K is similar to 
the value for the Curie temperature expected\cite{Jungwirth}
and observed elsewhere.\cite{Piazza}

As a starting point for a more detailed discussion, consider
an energy functional for the system. Introducing
coherent state creation operators $c_{\uparrow}^{\dagger}({\bf r})$
and $c_{\downarrow}^{\dagger}({\bf r)}$ for the two Landau levels involved,
correlations are characterised by the expectation value
of pseudospin, ${\bf S}({\bf r}) =
\langle c_{\alpha}^{\dagger} \mbox{\boldmath$\sigma$}_{\alpha \beta} 
c_{\beta} \rangle$, where \mbox{\boldmath$\sigma$} is the vector
of Pauli matrices. The order parameter has magnitude 
$|{\bf S}({\bf r})|=S$, where $S=1$
at a combined filling factor of unity for the two Landau levels
and is smaller otherwise.
For variations of ${\bf S}({\bf r})$
which are smooth on the scale of the cyclotron radius, one expects
the energy functional to have the form
\be
\label{energy}
E =\int \left(- D S_z^2 +  
J |\nabla {\bf S}|^2 +
\delta J\,|\partial_n {\bf S}({\bf r})|^2  
- h S_z 
\right) 
d^2{\bf r}\,.
\ee
Here $D>0$ represents Ising anisotropy, $J$ is the spin stiffness,
the derivative $\partial_n \equiv {\bf \hat n}\cdot \nabla$
acts in the direction of the in-plane magnetic field, denoted by $\bf \hat n$, 
and $\delta J$ represents spatial anisotropy in the spin
stiffness (for simplicity, we omit anisotropy in spin-space
from the stiffness). The effective
Zeeman field acting on pseudospins is $h$. In experiment, 
the strength of this field
varies through zero as the tilt angle $\theta$ of the sample
in the applied magnetic field is varied through the Landau level
coincidence point; its strength depends also
on  field magnitude $B_{\rm tot}$ and on carrier density $n$.

For a homogeneous system, the ground state of Eq.\,(\ref{energy})
is uniform with $S_z= {\rm sgn}(h) S$ and ${\bf S}_{\perp} = {\bf 0}$.
Domains may arise either in metastable states or because
they are induced by quenched disorder. While metastability and hysteresis
are observed in some examples of quantum Hall ferromagnets,\cite{Piazza}
this is not reported to be an important aspect of observations in
Refs.\,\onlinecite{Zeitler} and \onlinecite{Pan}.
We therefore turn to domains induced by disorder. Potentially the most
important source of disorder in Eq.\,(\ref{energy}) is
randomness in $h$, and behaviour
of the random field Ising model has been studied very 
extensively.\cite{Imry} It is useful to distinguish
the weak and strong disorder regimes: taking $h$ to fluctuate
about mean value zero with amplitude $\Delta$ and correlation
length $l$, and supposing $l$
is greater than the domain wall width $w=\sqrt{J/D}$, 
the boundary between the two regimes lies at
$l \Delta \sim \sqrt{J D}$. At weak disorder, domain size $\xi$ 
is much larger than $l$
and domain morphology depends, amongst other things, on the difference 
in energy per unit length of domain walls running
parallel or perpendicular to the in-plane magnetic field component:
this energy difference is of order $\sqrt{\delta J D}$.
At strong disorder, which we shall argue is the limit relevant here,
the domain pattern is simply that of ${\rm sgn}(h)$.

To apply these ideas, it is necessary to identify a
microscopic origin for such a random field. One
possibility is that variations in carrier
density $n$, arising either from impurity scattering or from large-scale
inhomogeneities, produce changes in the value of $h$. 
Randomness of this kind is expected to be spatially isotropic, but
may give rise to transport anisotropy via dependence
of the domain wall energy on spatial orientation. A second possibility is that
sample surface roughness changes the local value of $\theta$, 
and hence $h$. To compare the likely importance of these
two, we appeal to experiment, noting (for example, from
Fig.\,2 of Ref.\,\onlinecite{Zeitler}) that while the coincidence
transition has a rather small width ($0.5^{\circ}$) in
$\theta$, it has a much larger width ($20\%)$ in $B_{\rm tot}$,
which is indicative of its width in $n$.  
The existence of sample surface roughness of the required
amplitude, with $l\sim 1\mu{\rm m}$, is noted in Ref.\,\onlinecite{Zeitler}
and well-established in a variety of other contexts.\cite{roughness} 
Moreover, it can account for transport anisotropy, as we now
show. 

Let $z({\bf r})$ denote height of the sample surface above an
average reference plane, and let $\theta_c$ be the critical angle at which
the Landau level coincidence transition occurs. Then
for small-angle roughness
\be
h({\bf r}) = \alpha(\theta - \theta_c) + 
\alpha \partial_n z({\bf r})\,,
\ee
where $\alpha$
is a proportionality constant.
Crucially, by this mechanism surface roughness with a correlation
function 
\be
C({\bf r})=\langle z({\bf r'})z({\bf r + r'})\rangle 
- \langle z({\bf r'})\rangle \langle z({\bf r + r'})\rangle
\ee
which is {\it isotropic} generates a random field with {\it spatially
anisotropic correlations}, since
\be
\langle h({\bf r'})h({\bf r + r'})\rangle -
\langle h({\bf r'})\rangle \langle h({\bf r + r'})\rangle 
= - \partial_n^2 C({\bf r})\,.
\ee

\begin{figure}[ht]
\includegraphics[width=8cm]{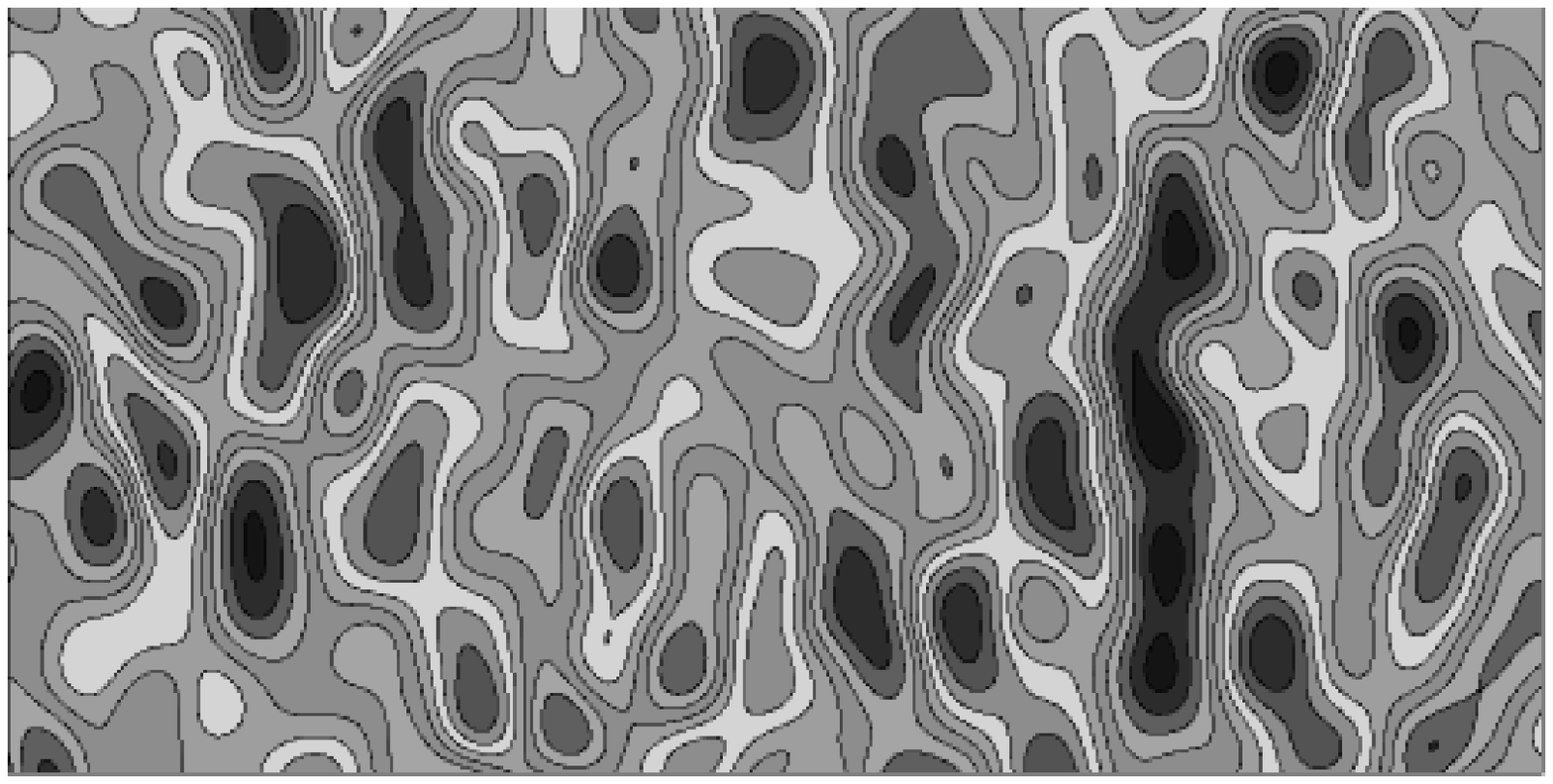} 
\vspace*{-12cm}

\includegraphics[width=8cm]{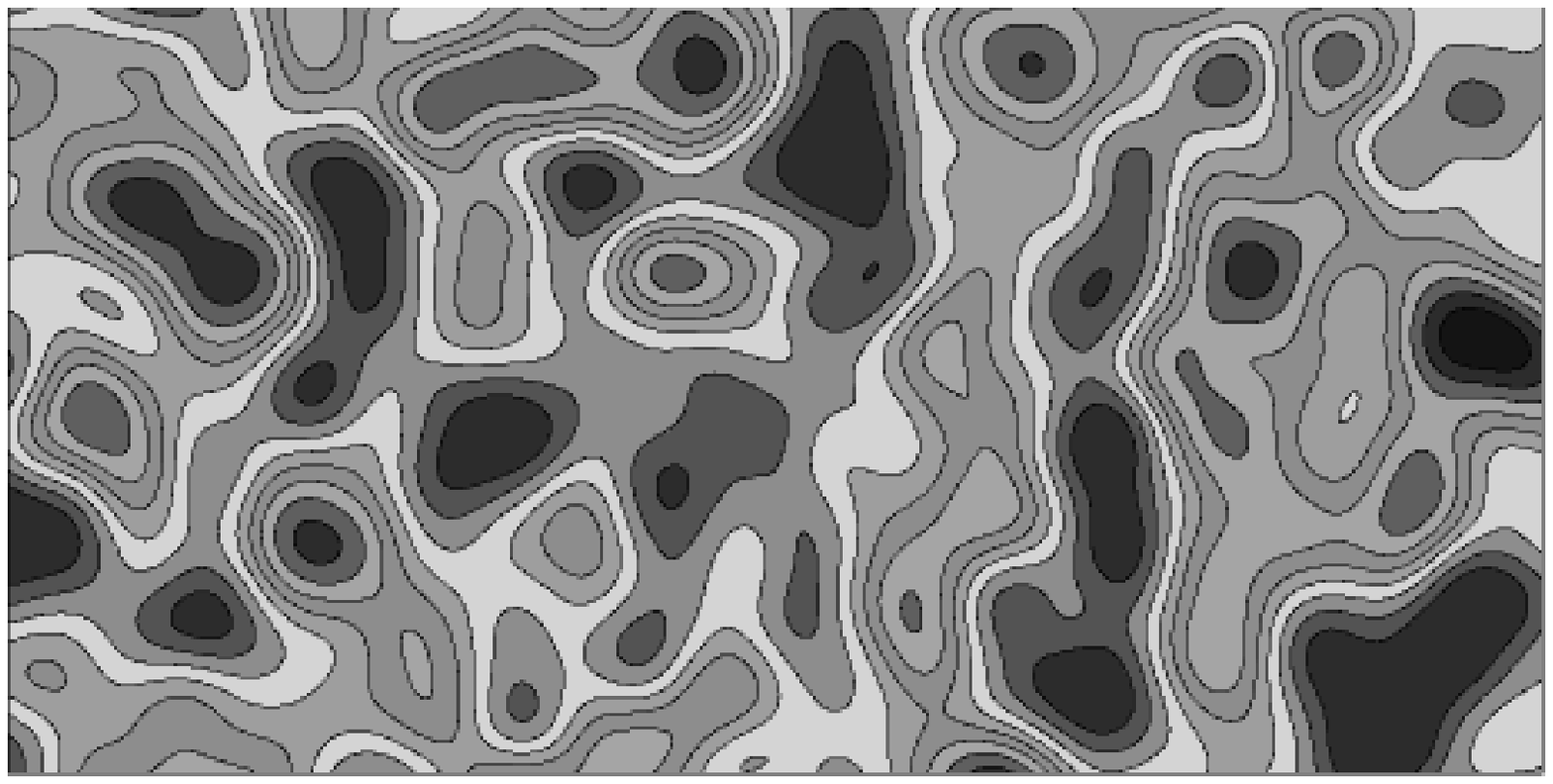}
\vspace*{3cm}

\caption{\label{snapshot} Greyscale plots of: (upper panel) spatially
isotropic surface roughness $z({\bf r})$, and (lower panel) the spatially
anisotropic Ising model random field $\partial_n z({\bf r})$ which
results from this surface roughness. 
}  
\vspace{-0.3cm}
\end{figure}

\begin{figure}[ht]
\includegraphics[width=8cm]{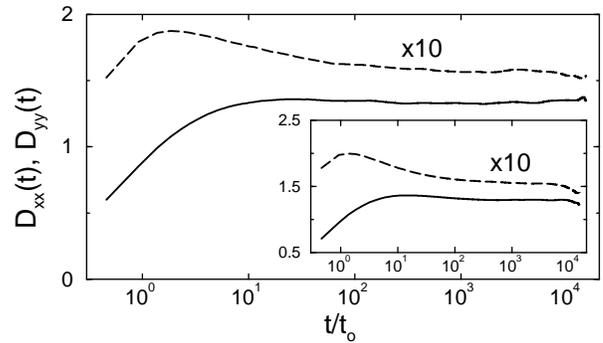}
\caption{\label{diffusion} Simulation data used to determine diffusion
coefficient anisotropy. Mean square displacements per unit time
in directions perpendicular  ($D_{xx}(t)$, full line)
and parallel ($D_{yy}(t)$, dashed line) to the in-plane field
direction $\bf \hat n$, averaged over all trajectories. Inset: averages over
open trajectories only, $\langle x^2(t) \rangle/t^{8/7}$ 
and $\langle y^2(t) \rangle/t^{8/7}$, demonstrating scaling with 
the classical percolation exponent value.  
}  
\vspace{-0.3cm}
\end{figure}

To establish the characteristic degree of this anisotropy, we have 
carried out numerical simulations. Taking $z({\bf r})$ to be a
superposition of overlapping Gaussian functions of position,
with centers placed at random points in the plane and
amplitudes distributed uniformly about zero, and setting
$\theta = \theta_c$, the resulting random field $h({\bf r})$ 
in a typical realisation is illustrated in Fig.\,\ref{snapshot}.
Anticipating our discussion of transport on domain walls,
we quantify the anisotropy evident in this figure by following
the classical dynamics of a particle that moves along contours 
of $h({\bf r})$, using methods described previously.\cite{Evers_old}
Averaging over randomly-placed starting points, 
we expect diffusive motion
in the sense that       
mean square displacements grow linearly in time.
Taking $\bf \hat n$ parallel to the $y$-axis,
the quantities
$D_{xx}(t)\equiv \langle x^2(t)\rangle/t$
and $D_{yy}(t)\equiv \langle y^2(t)\rangle/t$
should then approach the eigenvalues, $D_{xx}$ and $D_{yy}$,
of the diffusion tensor, for times $t$
which are large compared to the correlation time, $t_0$. 
Evidence that $D_{xx}(t)$ and $D_{yy}(t)$ indeed tend to a finite limit, 
with $D_{xx} \sim 8 D_{yy}$,
is presented in Fig.\,\ref{diffusion}.
The orientation of this anisotropy, with the larger
diffusion constant in the direction perpendicular to the
in-plane magnetic field component, is as observed in 
Refs.\,\onlinecite{Zeitler,Pan}, and its magnitude is about the
same as that determined at low temperature using a Hall
bar sample.\cite{Zeitler} The precise value of $D_{xx}/D_{yy}$
will be dependent on disorder distribution,
but we see no reason to expect large variations.
Our calculations also provide an opportunity to test
the universality class of our anisotropic percolation problem, since
diffusive growth in mean square displacement arises at 
long times from a balance between bounded motion on closed trajectories
and super-diffusive motion on trajectories which remain open
up to the observation time.\cite{Ziff} Averaging only over open
trajectories, one expects $\langle x^2(t)\rangle \propto 
\langle y^2(t)\rangle \propto t^{8/7}$, if the anisotropic
problem is in the same universality class as the standard classical
percolation problem: data shown in the inset to Fig.\,\ref{diffusion}
support this conclusion.

\begin{figure}[ht]
\includegraphics[width=6.5cm]{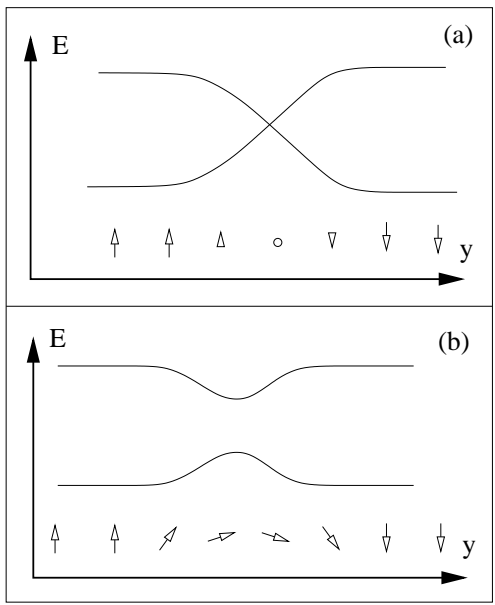} 
\caption{\label{wall} Schematic summary of domain wall structure,
showing pseudospin 
and excitation energies $E$, as a function
of position $y$ across the wall, within Hartree-Fock theory:
(a) for an Ising wall stabilised
by short-range scattering, and (b) for a
Bloch wall.
}  
\vspace{-0.3cm}
\end{figure}

The foregoing discussion is based on the idea that transport occurs
along boundaries between domains. In order to substantiate this,
we next examine transport properties of domain walls between
oppositely magnetised phases of the Ising quantum Hall ferromagnet.
Recalling that the domain wall forms the boundary between a region
on one side with filling factors for the coincident Landau levels of
$\nu_{\uparrow} \simeq 1$ and $\nu_{\downarrow} \simeq 0$,
and a region on the other with interchanged
filling factors,  $\nu_{\uparrow} \simeq 0$ and $\nu_{\downarrow} \simeq 1$,
the simplest structure one might imagine is that shown in 
Fig.\,\ref{wall}(a). In this picture, the wall supports 
two counter-propagating modes with opposite spin polarisation, which
arise as edge states of the occupied Landau levels in the domains
on either side. Such an Ising domain wall, in which 
${\bf S}_{\perp}({\bf r})={\bf 0}$ everywhere and ${\bf S}({\bf r})={\bf 0}$ at
the wall centre, may be stabilised by short-range scattering, which
allows solutions with $|{\bf S}({\bf r})|<1$,\cite{Yarlagadda}
in contrast to Eq.\,(\ref{energy}).
For a sample without short-range scattering, however,
Hartree-Fock theory yields\cite{Brey} the Bloch domain wall
structure shown in Fig.\,\ref{wall}(b). Here, 
${\bf S}_{\perp}({\bf r})\not={\bf 0}$ within the wall. In consequence,
within Hartree-Fock theory
there is mixing and an avoided crossing of 
edge states arising from occupied Landau levels on either
side of the wall. 

At this level of approximation,
for a Bloch wall the chemical potential lies within
a quasiparticle gap. To account for transport under these conditions, it
is necessary to consider collective excitations.
The combined consequences of continuous symmetry for 
the Hartree-Fock solution under rotations of 
$\langle {\bf S}_{\perp}({\bf r})\rangle$ about the Ising axis,
and the connection between spin or pseudospin and charge that is standard
for quantum Hall ferromagnets\cite{Sondhi} have been examined
in a related context in Ref.\,\onlinecite{Mitra}.
Introducing pseudospin rotation angle $\varphi$, 
as a function of position coordinate $x$ along the wall and 
imaginary time $\tau$, the action 
\be
S=\frac{\rho}{2}\int \int \left[ 
\left(\frac{\partial \varphi}{\partial x}\right)^2
+ \frac{1}{u^2}
\left(\frac{\partial \varphi}{\partial \tau}\right)^2
\right] dx\, d\tau
\ee
is obtained for domain wall excitations,\cite{Mitra}
where in our context $\rho \sim J w\sim e^2/\epsilon$ and 
$u\sim e^2/\epsilon \hbar$.
A charge density
$(2\pi)^{-1}\partial \varphi /\partial x$ is associated
with these modes. This is the action for a spinless Luttinger liquid.
A vital property for our argument is that left 
and right-moving excitations propagate independently, provided 
rotation symmetry about the pseudospin easy axis is exact. In short, 
Fig.\,\ref{wall}(a) remains a useful picture
even without short-range disorder to stabilise an Ising wall,
provided only that
there is no spin-orbit scattering. In this picture, transport in
a multidomain sample occurs at domain boundaries, via two independent, 
counter-propagating sets of modes. Neglecting quantum interference 
effects, we arrive at the problem for which numerical 
results are given above.

In conclusion, we have argued that the observations of anisotropic 
transport reported in Refs.\,\onlinecite{Zeitler} and \onlinecite{Pan}
can plausibly be attributed to formation of anisotropically shaped
domains, induced as a result of sample surface roughness.
Our numerical work demonstrates that this mechanism generates an 
anisotropy comparable to that found experimentally.\cite{Zeitler}
In addition, the onset temperature for strongly
anisotropic transport is comparable to the critical
temperature expected\cite{Jungwirth} and observed\cite{Piazza} 
in other Ising 
quantum Hall ferromagnets.
For future work it would be interesting to investigate 
transport in systems with deliberately induced
surface features.

We thank U. Zeitler for extensive discussions, and E. H. Rezayi
for correspondence. The work was supported
by  the A. v. Humboldt 
Foundation (J. T. C.) and by
the Schwerpunktprogramm ``Quanten-Hall-Systeme''
der Deutschen Forschungsgemeinschaft.

\end{document}